\documentclass[twocolumn, showpacs,preprintnumbers,amsmath,amssymb, superscriptaddress, prl]{revtex4}

\usepackage{graphicx}
\usepackage{dcolumn}
\usepackage{bm}

\begin{document}

\title{Even-order aberration cancellation in quantum interferometry}

\author{Cristian Bonato}
\affiliation{
Department of Electrical \& Computer Engineering, Boston University, Boston, Massachusetts 02215}

\affiliation{
CNR-INFM LUXOR, Department of Information Engineering, 
University of Padova, Padova (Italy)}

\author{Alexander V. Sergienko}

\affiliation{
Department of Electrical \& Computer Engineering, Boston University, Boston, Massachusetts 02215}

\affiliation{
Department of Physics, Boston University, Boston, Massachusetts 02215}

\author{Bahaa E.A. Saleh}
\affiliation{
Department of Electrical \& Computer Engineering, Boston University, Boston, Massachusetts 02215}

\author{Stefano Bonora}
\affiliation{
CNR-INFM LUXOR, Department of Information Engineering, 
University of Padova, Padova (Italy)}

\author{Paolo Villoresi}

\affiliation{
CNR-INFM LUXOR, Department of Information Engineering, 
University of Padova, Padova (Italy)}

\begin{abstract}
We report the first experimental demonstration of even-order aberration cancellation 
in quantum interferometry. The effect is a spatial counterpart of the spectral group velocity dispersion cancellation, which is associated with spectral entanglement. It is manifested in temporal interferometry by virtue of the multi-parameter spatial-spectral entanglement. Spatially-entangled photons, generated by spontaneous parametric down conversion, were subjected to spatial aberrations introduced by a deformable mirror that modulates the wavefront. We show that only odd-order spatial aberrations affect the quality of quantum interference.
\end{abstract}


\maketitle

The nonlinear optical effect of spontaneous parametric down conversion (SPDC) has been a reliable source of entangled-photon states for the last thirty years. A photon of the pump radiation has a random chance to be converted into two photons with lower energy while traveling though the nonlinear material.  Conservation of energy and momentum governs the spatial and spectral state of the down converted light. In case of a monochromatic pump beam, energy conservation leads to strong anticorrelation between the frequency components of signal and idler wave packets. This symmetric superposition of all possible anticorrelated frequencies with respect to the degenerate frequency of signal and idler waves gives rise to frequency entanglement. 

Even-order dispersion cancellation is among the most interesting consequences of frequency entanglement \cite{franson92, steinberg92a}. If one of the two photons of an entangled pair travels through a dispersive material and both photons are combined on a beamsplitter in a Hong-Ou-Mandel configuration \cite{hom87},  then the rate of coincidences between the counts of two single-photon detectors placed at the output ports depends on the odd-order dispersion terms only when observed as a function of the path difference between the two arms before the beamsplitter. The detrimental effect of even-order dispersion (such as group velocity dispersion), which leads to the wavepacket broadening, is cancelled. This has been exploited in several applications such as the measurement of photon traveling time trough a material \cite{steinberg92b}, and improving the accuracy of remote clock synchronization \cite{giovannettiPRL01}. Optical coherence tomography \cite {qoct02, nasr03} has also benefited from this nonclassical effect. This quantum effect has inspired recent developments of classical nonlinear optical systems mimicking dispersion cancellation \cite{resch07, shapiroOCT06}. 

 The wavevector of a monochromatic wave at a given frequency $\Omega$ has a bidimensional transverse wavevector $\mathbf{q}$ (in the plane orthogonal to the propagation direction) and a longitudinal component $\kappa (\mathbf{q}, \Omega) = \sqrt{\frac{n^2(\Omega)\Omega^2}{c^2} - |\mathbf{q}|^2}$. In parametric downconversion with a  plane-wave pump, momentum conservation leads to anticorrelation of the transverse wavevector components \cite{rubin96}.  This analogy with frequency anticorrelation \cite{dualityPRA00} suggests the existence of a spatial counterpart to dispersion-cancellation. However, no experimental observation of a spatial dispersion cancellation effect has been reported so far. 

The longitudinal component of the wavevector, on the other hand, sets up the phase-matching relation that establishes conditions for an effective energy conversion between three interacting waves, pump, signal, and idler. Since the longitudinal wavevector depends both on frequency and on transverse momentum, this condition sets a specific relation between the frequency and the emission angle of down converted radiation. In other words, the frequency and transverse momentum degrees of freedom cannot be factorized and the overall quantum state is concurrently entangled in both $\omega$ and $\mathbf{q}$ (multi-parameter entanglement). This leads to several interesting effects where the manipulation of a spatial variable affects the shape of the temporal interference pattern and also polarization interference pattern  \cite{mete04} .

In this Letter, we exploit the multi-parameter entangled states generated by SPDC to demonstrate the effect of even-order spatial aberration cancellation. We use an SPDC source to produce momentum-anticorrelated photons and we modulate their wavefronts by a transfer function $H (\mathbf{q})$. Due to the correlations between $\mathbf{q}$ and $\omega$, the manipulation in the $\mathbf{q}$-space will introduce changes in the $\omega$-space. Therefore the spatial wavefront modulation will affect the temporal interference pattern, which can be observed using a polarization two-photon interferometer\cite{exp_typeIIPRA94, mete04}. With this technique we show that, due to the anticorrelation of the transverse momenta, the even-order aberrations are cancelled-out and only the odd-order contributions influence the resulting interference pattern. We believe  this effect may lead to interesting applications in the field of quantum imaging.

The setup of our experiment is sketched in Fig. 1. A laser diode with a single longitudinal-mode selection (405 nm, 35 mW) pumps a 1.5-mm thick BBO crystal that is cut for a collinear degenerate type-II phase-matching. Approximating the pump beam with a plane-wave, the two orthogonally-polarized photons emitted by the crystal can be described by the quantum state \cite{rubin96} :
\begin{equation}
 \left | \psi \right \rangle = \int d\mathbf{q} \int d\omega \xi (\mathbf{q}, \omega) \hat{a}_H^{\dagger} (\mathbf{q}, \Omega_0 + \omega) \hat{a}_V^{\dagger} (-\mathbf{q}, \Omega_0-\omega) \left | 0 \right \rangle,
\end{equation}
where
\begin{equation}
\label {psi_bulk}
\xi (\mathbf{q}, \omega) =  \mbox {sinc} \left[ \frac {L \Delta(\mathbf{q}, \omega)}{2}\right] e^{i  \frac {L\Delta(\mathbf{q}, \omega)}{2}}. 
\end{equation}

The phase-mismatch function $\Delta(\mathbf{q}, \omega)$ in the paraxial approximation takes the form:
\begin{equation}
\label {Eq:delta}
\Delta(\mathbf{q}, \omega) = -\omega D + M \mathbf {\hat{e}_2}\cdot \mathbf{q}+\frac{2|\mathbf{q}|^2}{k_p} ,
\end{equation}
where D is the difference between the inverse of the group velocities of the ordinary and the extraordinary waves inside the birefringent crystal and M is their spatial walk-off in the vertical direction $\mathbf {\hat{e}_2}$. The term $\frac{2|q|^2}{k_p}$ is due to diffraction during the propagation of photons through the crystal. In case of a BBO crystal, phase-matched for a degenerate ($\lambda_0 = 810$ nm) collinear type-II downconversion, $D = 182$ ps/mm and $M = 0.0723$. 

A polarizing beamsplitter (PBS) separates the horizontally-polarized photon and the vertically-polarized one into two distinct paths, one towards a flat mirror (FM) and the other towards a deformable mirror (DM). Each photon passes through a 4-$f$ system comprising a lens (L1) of focal length $f = 200$ mm positioned at a distance $f$ from the output plane of the crystal, and the same distance $f$ from the mirror. On the way from the crystal to the mirror, the lens maps each wavevector component to a different point $ \mathbf{x} (\mathbf{q}, \omega)$ on the mirror surface. The limited downconversion bandwidth (about 30 nm for a collection angle of 25 mrad) allows us to neglect the frequency dependence of the 4-$f$ system, assuming that the lens is achromatic,
\begin{equation}
\mathbf{x} (\mathbf{q}, \omega) = \frac{f}{k(\omega)} \mathbf{q} \approx \frac{f}{k_o} \mathbf{q}.
\end{equation}
The deformation of the mirror surface at point $\mathbf{x}$ can be discribed by the function $\zeta (\mathbf{x})$. A photon focused to the point $\mathbf{x}$ by the lens will travel a distance $\zeta (\mathbf{x})$ to the mirror surface, will be reflected and will travel a distance $\zeta (\mathbf{x})$ back to the lens focal plane. Therefore it will acquire a phase shift:
\begin{equation}
 \varphi (\mathbf{x}) \approx 2 k_o \zeta (\mathbf{x})
\end{equation}
After reflection from the mirror, the same lens maps every point back to a wavevector. Mathematically, the transformation induced by the 4-$f$ system can be described by the transfer function:

\begin{equation}
H \left( \mathbf{q} \right)  = p \left( \frac{f}{k} \mathbf{q} \right)  e^{i \varphi (\frac{f}{k_o}  \mathbf{q})}, 
\end{equation}
where the pupil function $p (\mathbf{x})$ describes the circular aperture of the mirror.

The deformable mirror \cite{specchioRSI2006} consists of a thin nitrocellulose silver-coated membrane (12 mm diameter, 5 $ \mu \text{m} $ thick, initial flatness less than 20 nm rms) that is deformed by electrostatic forces created when a voltage drop (maximum 270 V) is applied to 37 electrodes. The action  produced by each actuator was mapped by measuring induced deformation with a Zygo interferometer, creating an influence function matrix. 

In addition, each photon, travelling from the PBS to the mirror and back  passes twice through a quarter-wave-plate oriented at 45 degrees  and flips its polarization state. This way the photon that has been transmitted is now reflected at the polarizing beamsplitter (PBS) and vice-versa, resulting in both photons leaving the modulation section together towards the polarization interferometer.

The polarization interferometer \cite{theory_typeIIPRA94, exp_typeIIPRA94} consists of a birefringent delay-line (DL, made of two sliding quartz wedges) providing a variable delay $\tau$ and a non-polarizing beamsplitter (BS) that splits the photons in two paths directed to two single-photon detectors $D_1$ and $D_2$. 
A polarizer oriented at 45 degrees is placed in front of each detector in order to erase information about the polarization of the incoming photon. Photons were collected by a lens in each arm and focused onto the detector's active area. To maximize the spatial collection capability we used two open-face (180 $\mu \text{m}$ diameter) single-photon silicon avalanche photodiodes. Using a fiber coupler would limit the number of collected spatial modes. All experiments have been performed with an 8-mm diameter pinhole placed at 330 mm from the output plane of the 4-$f$ system, therefore collecting photons from an angle of about 25 mrad. A dichroic mirror and a pair of interference bandpass filters with a bandwidth that is greater than that of downconversion have been used to reject the residual pump radiation and the background light. The number of coincidences acquired as a function of optical polarization delay $\tau$ shows a high-visibility quantum interference pattern in the form of a dip \cite{mete04}. 

\begin{figure} [ht]
\label {fig1:exp}
\centering
\includegraphics [width = 8 cm] {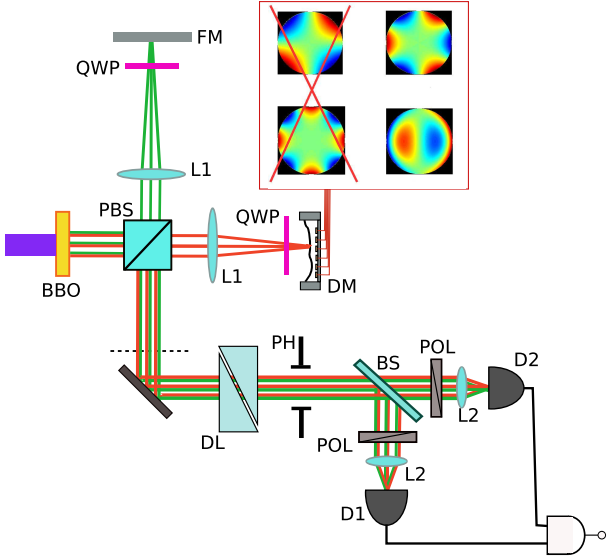}
\caption{Schematic of the experimental setup. Examples of aberrations induced by the deformable mirror are illustrated in the inset: the even-parity aberrations on the left are cancelled, while the odd-parity ones on the right alter the shape of the interference pattern.} 
\end{figure}

Since the photon-counting detectors are slow, compared with the coherence time of downconverted  photons, and since their surface is larger than the spot size, the expression for the coincidence rate, in the paraxial approximation, is \cite{bonatoPRA08}:
\begin {equation}
\label {Rc}
R_C (\tau)  = R_0 \left[ 1 - \Lambda \left( 1 - \frac {2\tau}{DL}\right) W_M (\tau) \right] ,
\end {equation}
where $R_0$ is the background coincidence rate,  $\Lambda (\alpha)$ is a triangular function [$\Lambda (\alpha)= 1 - |\alpha|$ if $|\alpha|<1,$ and $\Lambda (\alpha) = 0$, otherwise], and:

\begin{equation}
\label{xi_general}
\begin{split}
W_M (\tau) & = \int d\mathbf{q} d\mathbf {q'}e^{i \frac {2 d_1}{k_p} [|\mathbf{q}|^2 - |\mathbf{q'}|^2]} \tilde{P}_A \left[ \mathbf{q} + \mathbf{q'}\right] \times \\
& \qquad \mbox {sinc} \left[ ML \mathbf {\hat{e}_2}\cdot (\mathbf{q} + \mathbf{q'})\Lambda \left( 1 - \frac {2\tau}{DL}\right) \right] \times \\ 
& \qquad H \left(  \frac{f}{k}\mathbf{q}\right)  H^* \left( \frac{f}{k}\mathbf{q'}\right)  \quad e^{-i \frac {M}{D} \tau \mathbf {\hat{e}_2}\cdot (\mathbf{q} - \mathbf{q'})}.
\end{split}
\end{equation}

The function $\Lambda \left( 1 - \frac {2\tau}{DL}\right)$ represents a usual tringular dip one obtains in type-II quantum interferometry when working in the single spatial-mode approximation (using narrow pinholes). The function $W_M (\tau)$ takes into account the deformation of the triangular dip induced by the modulation of the transverse wavevectors and the Fourier transform of the shape of the detection apertures $\tilde{P}_A \left[ \mathbf{q} \right]$. In particular, the function $W_M (\tau)$ describes how manipulation in the $\mathbf{q}$ space by a filter $ H (\mathbf{q})$ is converted into a modification of the temporal interference pattern, by means of the coupling between wavevectors and frequencies set by the phase-matching conditions.

If the detection apertures are sufficiently large, the function $\tilde{P}_A \left[ \mathbf{q} + \mathbf{q'}\right]$ can be well approximated by a delta-function, so that Eq. (\ref{xi_general}) can be simplified to:

\begin{equation}
W_M^{R \rightarrow \infty} (\tau) = \int d\mathbf{q} \quad H^* \left( \frac{f}{k_0} \mathbf{q}\right) H \left( -\frac{f}{k_0}\mathbf{q}\right)  \quad e^{i \frac{2Mk_0}{fD}\tau \mathbf {\hat{e}_2}\cdot\mathbf{q}}.
\end{equation}

If the function $H (\mathbf{q})$ has a circular symmetry, its phase $\varphi (\mathbf{q})=  \arg{\{H (\mathbf{q}) \}}$ can be expanded using a set of Zernike polynomials, which are orthogonal on the unit circle \cite{BornWolf},
\begin{equation}
\label {LA_zernike}
\varphi (\mathbf{q}) = \sum_{n} \sum_{m} \varphi_{nm} R_n^m (\rho)\cos (m\theta),
\end{equation}
where $\mathbf{q} = (\rho \cos\theta, \rho \sin\theta)$ and $m = -n, -n+2, -n+4...n$.
Using the fact that $-\mathbf{q} = [\rho \cos (\theta+\pi), \rho \sin (\theta + \pi)]$, and that if $m$ is even, then $\cos [m(\theta+\pi)] = \cos (m\theta)$, while if $m$ is odd, then $\cos [m(\theta+\pi)] = -\cos (m\theta)$, one gets:
\begin{equation}
\label {LA_dispCanc}
\varphi (\mathbf{q}) - \varphi (-\mathbf{q}) = 2 \sum_n \sum_{m \quad odd} \varphi_{nm} R_n^m (\rho) \cos (m\theta).
\end{equation}

This means that only Zernike polynomials with $m$ odd (and consequently $n$ odd) will contribute to the shape of the interference pattern. For example, contributions from astigmatism ($n = 2, m = \pm 2)$, defocus ($n = 2, m = 0$), and spherical aberration ($n = 4, m = 0$) will all be cancelled.  On the contrary, coma ($n=3, m = \pm1$) and trefoil ($n = 3, m= \pm3$) will be present.

\begin{figure} [ht]
\label {fig2:coma}
\centering
\includegraphics [width = 9 cm] {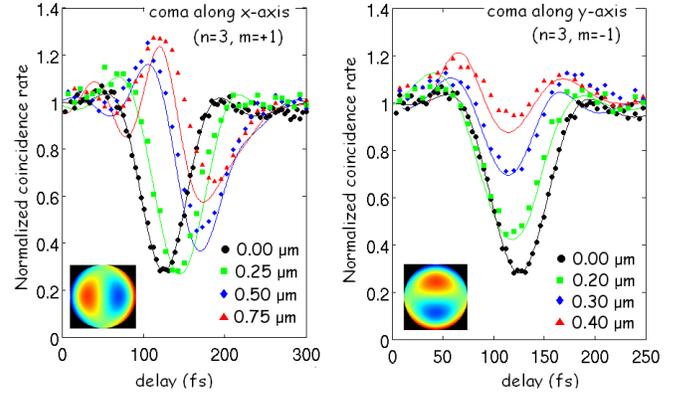}
\caption{Coincidence-rate interference patterns when coma ($n=3, m = \pm1$) along the x-axis (on the left) and along the y-axis (on the right) is imposed on the deformable mirror. Solid lines illustrate theoretical fitting with experimental parameters. The initial relative tilt between the deformable mirror (DM) and the flat mirror (FM) is used as an adjustable parameter to account for the imperfectness of experimental alignment between two arms. The shapes of adaptive mirror deformation producing selected aberrations are illustrated in the insets.}
\end{figure}

\begin{figure} [ht]
\label {fig3:astigmatismo}
\centering
\includegraphics [width = 9 cm] {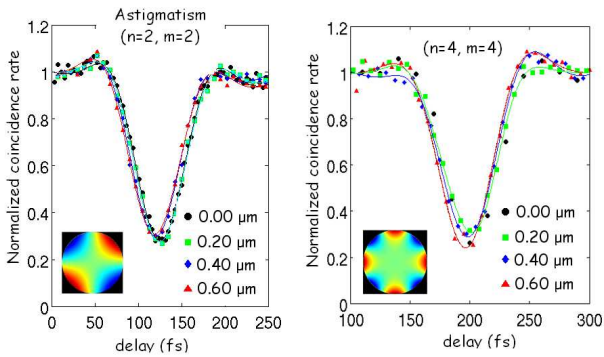}
\caption{Interference patterns in the coincidence rate when astigmatism at 45 degrees (left picture) and the aberration described by $n = 4, m = 4$ are dialed on the mirror. The intensity of the aberrations has been scanned from $0.2 \mu \text{m}$ to $0.8 \mu \text{m}$ peak-to-valley deformation. The interference pattern is insensitive to the aberrations within experimental errors.}
\end{figure}

We studied different types of aberrations: coma along the x-axis ($n = 3, m = +1$), coma along the y-axis ($n = 3, m = -1$), astigmatism ($n = 2, m = 2$) and the aberration correponding to $n = 4$ and $m = 4$.  The experimental data for coma oriented along the $x$ and $y$ (parallel to the vertical-polarization) directions are presented in Fig. 2. Due to the multi-parameter entanglement of the two-photon state,the wavefront distortion induced by the deformable mirror modulates the spectral degree of freedom, rsulting in a modification of the temporal interference pattern. We increased the maximum amplitude of the mirror deformation from 0.2 $\mu \text{m}$  ($\sim 0.25 \lambda_0$) to 0.75 $\mu \text{m}$  ($\sim \lambda_0$) (peak-to valley deformation) for coma along $x$ and from 0.2 $\mu \text{m}$ to 0.4 $\mu \text{m}$  ($\sim 0.5 \lambda_0$) for coma along y axis. The shape of the interference pattern is changed dramatically when the intensity of the coma aberration increases. Theroretical predictions (solid lines) based on the experimental parameters are superposed to experimental data in Fig. 2: the matching between the two curves is pretty good. 

Experimentally obtained data for astigmatism (with symmetry axes oriented at 45 degrees with respect to the $x-y$ axes) and for the aberration identified by $n = 4$ and $m = 4$ are shown in Fig. 3. The effect of these even-order aberrations is effectively cancelled out due to the spatial correlations between the photons in parametric down conversion. Therefore, such type of spatial aberrations do not affect the shape of the dip so that the known quantum interference pattern is retained.

This effect has a clear analogy with even-order frequency dispersion cancellation due to frequency entanglement in SPDC.  In case of spectral dispersion, the use of a non-monochromatic pump reduces the degree of correlation between spectral components of entangeld photons and degrades the dispersion cancellation effect. In our case of even-order aberration cancellation, sharp spatial correlations can be obtained only in the approximation of a plane-wave pump beam. Therefore, wavevector correlations get weaker for focused pump beams and the aberration cancellation effect also degrade when the pump beam is focused tightly on the crystal. Furthermore, the spectral dispersion cancellation effect works in the limit of slow detectors because it requires integration over the temporal degree of freedom. Similarly, the even-order aberration cancellation works well only in the situation when the collection apertures used in experiment are sufficiently large to enable effective integration over the spatial degrees of freedom \cite{bonatoPRA08}.

The question remains whether the effect we have reported is purely quantum, or some classical counterpart can be envisioned, as in the case of spectral dispersion cancellation. We believe, that a classical optical configuration mimicking even-order aberration cancellation could potentially be discovered by exploiting a light source with strong degree of spatial intensity correlation similar to optical speckles.

In conclusion, we have demonstrated experimentally the effect of cancelling even-order spatial aberration in quantum interferometry using entangled-photon states generated in a type-II spontaneous parametric downconversion process. This effect may prove helpful in enhancing the spatial resolution in quantum imaging.

This work was supported by a U. S. Army Research Office (ARO) Multidisciplinary University Research Initiative (MURI) Grant; by the Bernard M. Gordon Center for Subsurface Sensing and Imaging Systems (CenSSIS), an NSF Engineering Research Center; by the Intelligence Advanced Research Projects Activity (IARPA), ARO through Grant No. W911NF-07-1-0629 and DEI-UniPD QUINTET strategic program. C. B. also acknowledges financial support from Fondazione Cassa di Risparmio di Padova e Rovigo.

\end{document}